\documentclass[prl,twocolumn,nopacs,preprintnumbers,notitlepage,amsmath,amssymb,superscriptaddress]{revtex4-2}

\usepackage{graphicx}
\usepackage[utf8]{inputenc}
\usepackage{xcolor}
\RequirePackage[normalem]{ulem} 
\RequirePackage{color}\definecolor{RED}{rgb}{1,0,0}\definecolor{BLUE}{rgb}{0,0,1} 

\begin{document}

\title{Verifying the magnitude dependence in earthquake occurrence}

\author{Giuseppe Petrillo}
\affiliation{The Institute of Statistical Mathematics, Research Organization of Information and Systems, Tokyo, Japan}
\author{Jiancang Zhuang}
\affiliation{The Institute of Statistical Mathematics, Research Organization of Information and Systems, Tokyo, Japan}

\date{\today}
\begin{abstract}
The existence of magnitude dependence in earthquake triggering has been reported. Such a correlation is linked to the issue of seismic predictability and remains under intense debate whether it is physical or is caused by incomplete data due to short-term aftershocks missing. Working firstly with a synthetic catalogue generated by a numerical model that capture most statistical features of earthquakes and then with an high-resolution earthquake catalogue for the Amatrice-Norcia (2016) sequence in Italy, where for the latter case we employ the stochastic declustering method to reconstruct the family tree among seismic events and limit our analysis to events above the magnitude of completeness, we found that the hypothesis of magnitude correlation can be rejected.

\end{abstract}
\maketitle

\emph{Introduction--}
The question of whether earthquakes can be predicted is one of the most important in both social and scientific contexts \cite{GJKM97, kagan97}. The study of earthquake occurrence phenomena is of great interest and involves multiple fields of research and technology, including engineering, geophysics, seismology, statistical mechanics, and more. It has been well known that seismicity is not completely random and the biggest predictable component in seismicity is clustering.  The Epidemic Type Aftershock Sequence (ETAS) model is considered as the standard baseline for modelling earthquake clusters \cite{ogata1988, ogata1998, PZ22, zhuang2011, zhuang2012} and short-term aftershock forecasting. In the traditional ETAS model, all the event magnitudes are assumed to be independent from the occurrence times and identically from the same random distribution -- the Gutenberg-Richter law, which in fact implies the complete randomness of earthquake magnitude in predictability. 

Recently, some researchers reported  the presence of correlations between seismic magnitudes within an earthquake sequence \cite{PZ22c,NOS19,NOS22,SS16,LdAG08}, i.e., subsequent events tend to have larger magnitudes than expected based on the Gutenberg-Richter law.  This implies  that there is some predictability from complete randomness for forecast earthquake magnitude since it is possible to predict in some extend the magnitude of an earthquake from a seismic signal before its rupture process completes. 

However, there also have been argues that such appear correlation is caused by  the short-term aftershock incompleteness (STAI) \cite{PZ22c,DG11}, which refers to the lack of recorded earthquakes following a major event due to overlapping coda-waves, particularly in the immediate aftermath of a large earthquake \cite{LCGPK16, hainzl16a, hainzl16b, dAGL18, LGMGdA17, LGdA19}. STAI does not only lead to a bias in the estimation of model parameters and in forecasting but also creates the apparent magnitude correlation.

Both STAI and magnitude correlation seem to be able to explain each other.  The existence of magnitude dependence offers an alternative explanation for STAI. In other words, the lack of recorded earthquakes following a major event may not be a recording issue but rather a preference to trigger earthquakes of a certain magnitude. It is important to note that the incompleteness of the instrumental seismic catalogue due to the overlapping of coda-waves is a well-established effect, and the supporters of the existence of correlations between magnitudes do not deny the existence of STAI. Rather, they attribute the absence of minor events to both instrumental issues and physical phenomena caused by magnitude clustering.

The traditional ETAS model does not account for either STAI or magnitude dependence. 
To improve the ETAS model's ability to describe seismicity, we need to make a choice between two different approaches. The first is to tackle the influence of artificial incompleteness,  by ``obscuring" events produced by a simulated ETAS catalog to reproduce the correct sequence of events present in the real catalog (as suggested in \cite{dAGL18,hainzl16b,OK06}) or by ``reconstructing" the complete catalog by reintroducing missing events (as suggested in \cite{ZWK20,SF21}). The second approach introduces a ``constrained" magnitude frequency distribution $P(m\mid m^*)$ for the aftershocks that are triggered directly by a parent event magnitude $m^*$ in the ETAS model to account for the existence of correlations between seismic magnitudes. Both approaches appear to improve the ETAS model's ability to describe seismicity, but it is still unclear which one corresponds to the reality and to be used for the next generation of seismic forecasting statistical models. 

In this article, we study the magnitude correlations for a synthetic seismic catalogue produced with a 2-layer OFC model (\cite{PLLR20}) which is able to produce a realistic earthquake statistics. Therefore, we propose a direct correlation analysis of a machine learning high-resolution catalogue for the Amatrice-Norcia (2016) sequence in Italy while avoiding biases due to uncertainty about descendants in the triggering phase. In fact, using stochastic declustering technique \cite{ZOV02,ZOV04}, we assign to each event $j$ a probability of being the offspring of a previous event $i$, or a background event. By declustering the instrumental catalog, we calculate correlations weighting the results based on the probability of the two events being correlated. After establishing the completeness magnitude of the catalog and performing statistical analysis on correlated pairs, we can check whether the magnitude correlation hypothesis can be rejected with an high level of confidence. \\

\emph{The Physical Model and Magnitude Correlation}
We implement the model defined in \cite{LPLR19,PLLR20,LPLR21} and tested in \cite{PRL22} composed by two elastic layers. The first one represents the brittle fault, the second, instead, is ductile. The aftershocks on the fault are nucleated by the interaction with the second layer. We consider a rectangular fault modeled as a lattice of blocks of size $L_x = 1000$ and $L_y = 400$. The stress acting on the $i$-th block is the sum of two contributions which take into account for the intra-layer and inter-layer interaction. The friction in the two layers is different, being velocity weakening (modelled as Coulomb Failure Criterion) in the brittle layer and velocity strengthening in the ductile layer. The ingredients introduced for friction induce stick-slip dynamics in the ductile layer, which allows all the statistical laws of earthquakes to be recovered. For more details on the model, see reference \cite{PLLR20}. The output seismic catalogue we use in this study contains $\sim 5,000,000$ events. \\ 

\begin{figure}[ht!]  
    \centering
    \includegraphics[width=9cm]{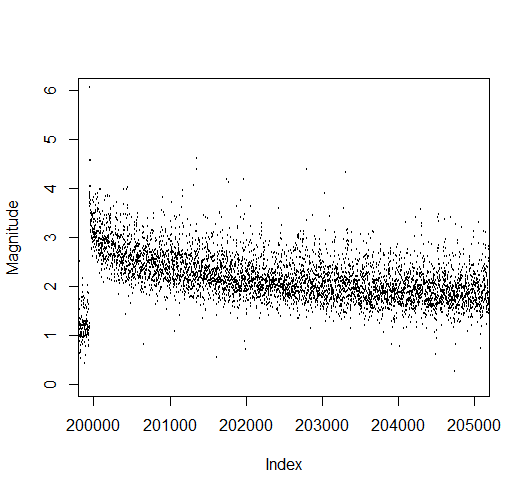}
    \caption {A slice of the Italian seismic catalogue plotting the magnitude versus the index of the event. The effect of short-term incompleteness is clearly recognisable.}
\label{incompletezza}
\end{figure}

\begin{figure}[ht!]  
    \centering
    \includegraphics[width=9cm]{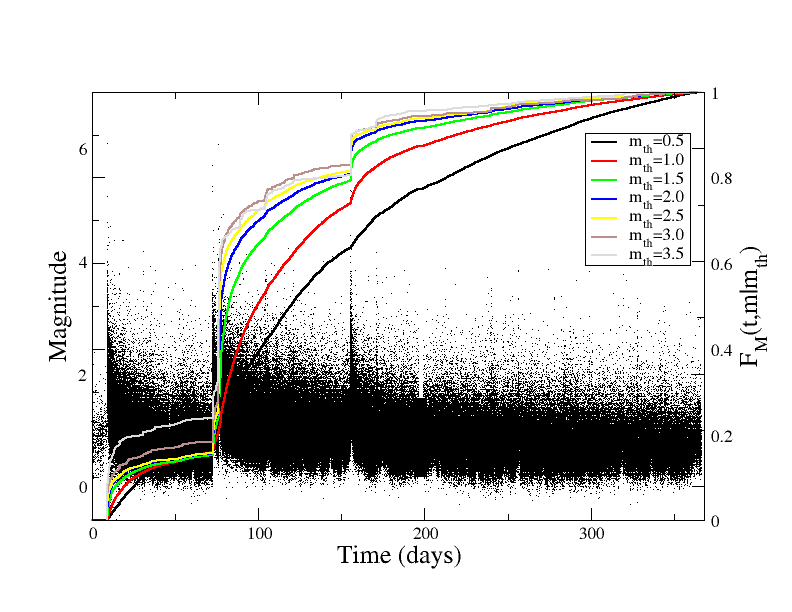}
    \caption{The Amatrice-Norcia catalogue (black dots) and the related $F_M (t,m|m_{th})$ function versus the time (coloured lines). Different colour represents a different choice of $m_{th}$ (see the legend). }
\label{compl}
\end{figure}

\emph{Completeness of the Amatrice-Norcia seismic catalogue--} 
The Machine‐Learning‐Based High‐Resolution Earthquake Catalog consists in $885,616$ events spanning a 1 year period, based on arrival times derived using a deep‐neural‐network‐based picker \cite{JFWMWMCGM21}. It is well known that immediately after a large earthquake, many aftershocks cannot be recorded (Fig.\ref{incompletezza}). The seismic waveforms generated by the aftershocks, many of which occur shortly after the mainshock, overlap with each other and cannot be accurately distinguished. Therefore, catalog completeness, is quantified in terms of a minimum threshold $m_c$ defined as the magnitude above which all events are identified and included in the seismic catalogue. The value of $m_c$ depends on the level of noise present in the seismic data and on the distance between the earthquake epicenter and the recording seismic stations \cite{MWWCW11}. Several methods have been proposed to estimate $m_c$ \cite{WW00,CG02,OK07,godano2017,GLP23,GP23} but many of these have limitations. To address the problem of calculating $m_c$, we estimate the completeness magnitude of the catalogue by plotting the quantity 
\begin{equation}
    F_M(t,m|m_{th})=\sum_{i=1}^N \frac{\mathbf{1}(t<t_i,m<m_{th})}{\mathbf{1}(m<m_{th})}
\end{equation}
where $\mathbf{1}$ is the indicator function and $m_{th}$ is the threshold magnitude chosen for the calculation of the quantity. In Fig.(\ref{compl}) it is easy to see how that for small values of $m_c$, the curves are distinctly separate, whereas they blur for larger values ($m_c \geq 2$). Neglecting the small noise, a complete collapse of all curves means that the catalog is complete and all occurred events have been recorded. Here, we consider complete the catalogue considering only earthquakes with $m>3$. 

\emph{Stochastic declustering--}
The main weakness of a direct statistical approach in calculating the correlations between magnitudes is that the calculation is performed by ordering the earthquakes chronologically and for a fixed spatial region. Thus there is a non-zero probability that related events occurring close-in-time are spatially distant. Conversely related events occurring close-in-space can be separated by a very large time interval. For this reason a simple space-time window selection is not suitable for this kind of study. To overcome this problem we employ the stochastic declustering methodology introduced by \cite{ZOV02}, with which is possible to estimate the probability that an event is a spontaneous event or is instead triggered by others. We define as $\rho_{ij}$ the probability that an event $i$ is an offspring of an event $j$. Since we are only interested in understanding whether there is magnitude clustering between the triggering events, we remove all background events from the computation, i.e. all events having $\rho_{ij}$ with $i=j$. After the procedure, we obtain a probability tree among the events. In particular, we built a matrix $(i,j,\rho_{ij})$, where $j$ is the possible mother of $i$, ranging from $1$ up to the total number of mothers, while $i$ is the index of the possible offspring related to $j$ ranging from $0$ up to the total number offsprings. We obtain $N_c=706,266$ combination of events with magnitude $m \geq 3$.

\begin{figure}[ht!]  
    \centering
    \includegraphics[width=8cm]{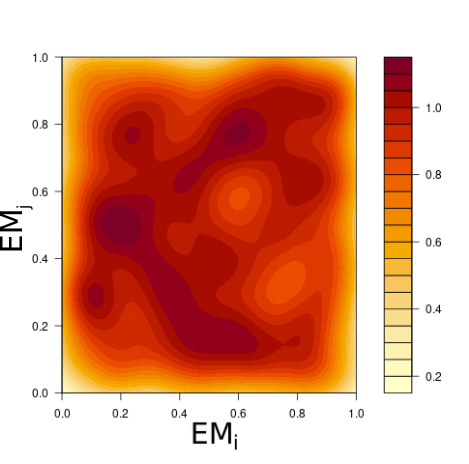}
    \caption{A biscale empirical transformation (BEPIT) of the quantities $ecdf(m_i)$ and $ecdf(m_j)$. The points are distributed randomly in the square $[0,1]\times[0,1]$ and no pattern is recognizable. }
\label{square}
\end{figure}

\begin{figure}[ht!]  
    \centering
    \includegraphics[width=8.5cm]{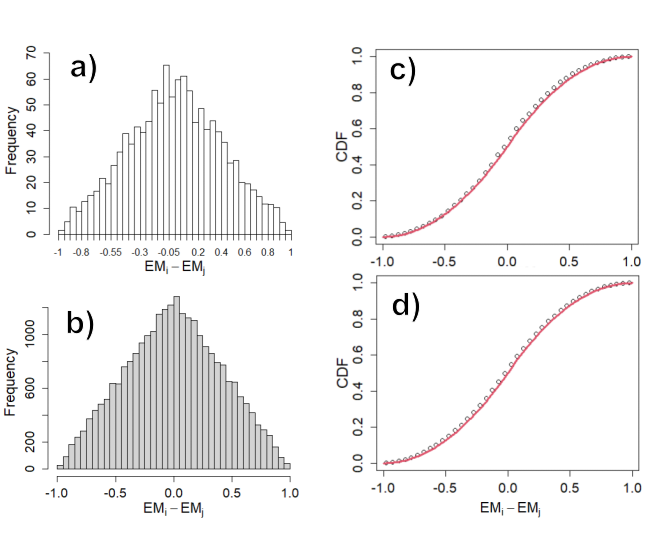} 
    \caption{The probability density function (pdf) of the quantity $\Delta_{ij}=EM_i-EM_j$ for the instrumental catalogue (a) and for the numerical catalogue (b). The triangular shape of the distribution suggests independence between the magnitudes. The cumulative density function (cdf) of $\Delta_{ij} = EM_i-EM_j$ (black empty circles) and the theoretical independent cdf (solid red line) for the instrumental catalogue (c) and for the numerical catalogue (d).}
\label{histo}
\end{figure}

\emph{Correlations of the Empirical Magnitudes--}
Instead of looking at the pairs $(m_i,m_j)$ directly, we estimate the counts 
$(EM_i, EM_{j})=(ecdf_{m_{1:n}}(m_i),  ecdf_{m_{1:n}}(m_{j}))$ in the unit square $ [0,1] \times [0,1]$ on a regular grid weighted considering the probability $\rho_{ij}$ (see Supp. Mat.). 
If there is no magnitude dependence, these points are distributed completely homogeneously in the square unit without any regular patterns (Fig.(\ref{square})). To statistically test whether a correlation exists, we can compute a $\rho_{ij}$-weighted histogram of the differences between $EM$-values, $\Delta_{ij} = EM_i-EM_j$ (Fig.(\ref{histo}a,\ref{histo}b)). For the null hypothesis of no-correlation, $\Delta_i$ has a probability density function (pdf) with a triangular shape:  $\Delta_i +1$ if $-1 < \Delta_i < 0$ , and $1-\Delta_i$ if $0< \Delta_i < 1$ (see Suppl.Mat.). Conversely, if $m_i$  and $m_{j}$ are positively correlated, then the pdf of $\Delta_{ij}$ will be more concentrated around $0$. In Fig.(\ref{histo}c,\ref{histo}d) the cumulative density function (cdf) of $\Delta_i$ is compared with the theoretical one for the null hypothesis. We find that the hypothesis of magnitude dependence is rejected for $m \geq 3$, conversely, for $m < 3$ a concentration of points around $0$ is more evident and the hypothesis of magnitude correlation cannot be rejected. 
We then justify the observed correlations observed if we consider all the events in the catalogue as "spurious" and caused to the lack of events with minor magnitudes not present in the catalogue. In conclusion we stat that the magnitude dependence we found in machine learning Amatrice-Norcia catalogue might be due to the short-term aftershock missing and cannot be attributed to a real dependence between magnitudes.
   
\emph{Conclusions--}
Resolving the magnitude correlation debate is crucial to focus statistical seismologists on developing a next-generation epidemic model. Moreover, the presence of correlations is also intrinsically linked to greater predictability of a seismic event. In this article we have shown how the correlation between magnitudes is an artificial effect due to the incompleteness of the instrumental catalog caused in turn by the overlapping of coda-waves. In addition to what has been performed in the literature, we propose three improvements: 1) we study the correlations on a synthetic catalogue produced by a physical model that captures the real statistical features of earthquakes, 2) we use a high-resolution experimental machine learning catalogue, 3) to be sure of calculating the correlation between the right pairs of events (father and descendants), we use the technique of stochastic declustering. \\
We want to underline that the proposed ETAS models with magnitude correlation may still perform well, however, it is likely that they do not capture the real process behind it. \\

\acknowledgments{This research activity has been supported by MEXT Project for Seismology TowArd Research innovation with Data of Earthquake (STAR-E Project), Grant Number: JPJ010217. We would like to acknowledge David Marsan for the useful discussion}.

\bibliography{magcorrprl2}

\end{document}